\documentclass[conference]{IEEEtran}
\IEEEoverridecommandlockouts

\def\re{\,{\rm Re}}
\def\im{\,{\rm Im}}

\def\bmat#1{\left[ \begin{array}{#1}}
\def\emat{\end{array} \right]}
\def\bdet#1{\left| \begin{array}{#1}}
\def\edet{\end{array} \right|}

\def\bbox#1{\textbf{#1}}

\def\cxi{\bbox{i}}

\def\cxv{\bbox{v}}

\DeclareFontFamily{U}{omega}{}
\DeclareFontShape{U}{omega}{m}{n}{<5->omsegr}{}%
\DeclareFontShape{U}{omega}{b}{n}{<5->omsegrb}{}%
\DeclareFontShape{U}{omega}{m}{it}{<5->omsegri}{}%
\DeclareFontShape{U}{omega}{b}{it}{<5->omsegrbi}{}%
\DeclareSymbolFont{Omeb}{U}{omega}{b}{n}
\DeclareMathSymbol{\cxpsi}{\mathord}{Omeb}{"79}
\DeclareMathSymbol{\cxPsi}{\mathord}{Omeb}{"59}

\def\h1{\hspace{10mm}}

\usepackage{amsmath,amssymb}
\usepackage{float}
\usepackage{lscape}

\usepackage{microtype}
\usepackage{siunitx}
\usepackage{algorithm,algpseudocode}

\usepackage{multicol,multirow}
\usepackage{amsthm}
\usepackage{xparse}
\usepackage{graphicx}
\usepackage{tikz}
\usetikzlibrary{shapes,arrows, matrix, calc}
\usepackage{threeparttable}
\usepackage{comment}
\usepackage{makecell}
\usepackage{colortbl}
\usepackage{hyperref}
\usepackage{subcaption}
\def\BibTeX{{\rm B\kern-.05em{\sc i\kern-.025em b}\kern-.08em
		T\kern-.1667em\lower.7ex\hbox{E}\kern-.125emX}}

\usepackage{tabularx}
\usepackage{pgfplots}
\usetikzlibrary{calc,patterns,angles,quotes}
\usepackage{mathtools}
\usepackage{xfrac}
\usepackage{bm}
\usepackage{cite}

\pgfplotsset{compat=1.18}

\graphicspath{{./figures/}}

\graphicspath{{./figures/}}

\newcounter{problemC}

\newcommand{\dc}{\mathrm{dc}}

\newcommand{\ab}{\alpha\beta}

\newcommand{\dq}{\mathrm{dq}}
\newcommand{\dcom}{\mathrm{d}}
\newcommand{\qcom}{\mathrm{q}}

\newcommand{\pu}{\mathrm{pu}}
\newcommand{\pcc}{\mathrm{pcc}}

\newcommand{\rf}{\mathrm{ref}}
\newcommand{\vir}{\mathrm{virt}}

\newcommand{\ext}{\mathrm{ext}}

\newcommand{\rr}{\mathrm{r}}

\newcommand{\mpppc}{MP$^3$C}

\allowdisplaybreaks

\DeclareSIUnit \pu {pu}
\DeclareSIUnit \voltampere {VA}

\pgfplotsset{compat=1.18}

\begin{document}

\title{Robust black start of an offshore wind farm with DRU based HVDC link using power synchronization control\looseness=-1
}

\author{Orcun Karaca, Ioannis Tsoumas, Mario Schweizer, Ognjen Stanojev, Lennart Harnefors
	\thanks{O. Karaca, M. Schweizer, and O. Stanojev are with ABB Corporate Research, Switzerland. emails: {\tt \{orcun.karaca, mario.schweizer, ognjen.stanojev\}@ch.abb.com}. I. Tsoumas is with ABB System Drives, Switzerland. email: {\tt ioannis.tsoumas@ch.abb.com}. L. Harnefors is with ABB Corporate Research, Sweden. email: {\tt lennart.harnefors@se.abb.com.} }}

\maketitle

\begin{abstract}
	This paper introduces a universal power synchronization controller for grid-side control of the wind turbine conversion systems in an offshore wind farm with a diode rectifier in the offshore substation of the HVDC link. The controller incorporates voltage-power droop controllers in the outer loop to enable the operation of this setup. To effectively handle the impact of large delays during black start and power ramp phases, virtual active and reactive power quantities are defined. These quantities are computed based on the current references prior to any modifications that might be needed to meet converter current and voltage limits or source constraints. Utilizing them in the outer loop ensures a balanced power sharing and a stable operation whenever the original (unmodified) current references are not realized. Case studies confirm the robustness of the proposed controller.\looseness=-1
\end{abstract}


\section{Introduction}

Among renewable energy sources, the global capacity of offshore wind farms (OWFs) has been rapidly increasing thanks to substantial investments and recent technological advancements. An OWF with a diode rectifier unit (DRU) based HVDC link connection offers several advantages over the one with a traditional voltage source converter-based HVDC link connection. These advantages include the reduced offshore substation complexity, reduced size and price as well as the increased reliability and efficiency. Consequently, there has been significant interest in research for such OWFs. The EU-funded project PROMOTioN\footnote{\href{http://www.promotion-offshore.net}{promotion-offshore.net}} delivered a proof of concept for the operation of a DRU-based HVDC link in combination with grid-forming wind converters. More recently, advanced grid-forming concepts for a similar setup have been developed within the $\text{FlexH}_2$ project\footnote{\href{http://www.grow-flexh2.nl}{grow-flexh2.nl}}, in which ABB is contributing. This project aims to design a novel OWF that also incorporates an onshore hydrogen production. 

Grid-forming control is becoming essential for the stable and reliable operation of modern power systems, particularly as we integrate more renewable energy sources with power electronic interfaces. This control strategy is characterized by its ability to support islanded operation, enabling converters to establish and regulate both voltage and frequency. In OWFs with a DRU-based HVDC link connection, as considered in this work, grid-forming control is not only desirable but also indispensable for the grid-side control of wind turbine (WT) converters. This necessity arises from the fact that a DRU cannot form the voltage and lacks the ability to control power flow actively. 

In order to enable this type of OWF configuration, several grid-forming control proposals have been made for the grid-side control of WT converters~\cite{blasco2010distributed,bernal2012efficiency,cardiel2018decentralized,asensio2019reactive,tang2021dru,zhang2022grid,yu2019analysis,bidadfar2020control}. Among these works, many, such as \cite{blasco2010distributed,bernal2012efficiency,cardiel2018decentralized,asensio2019reactive}, rely on reactive power synchronization (or $Qf$-droop), motivated also by the dynamics of the large capacitor bank for reactive power compensation. The authors in \cite{yu2019analysis} introduced an additional coupling from active power to angle via the so-called angle feedforward control to better dampen oscillations observed between the WTs. As an alternative,~\cite{bidadfar2020control} employs a $QV$ droop with additional couplings from frequency to voltage magnitude, and from active power to angle in a phase-locked loop (PLL)-based scheme, again better mitigating reactive power swings among WTs. 

In contrast to these existing works, our control design utilizes the universal framework proposed in~\cite{harnefors2020universal}. 
This architecture integrates the well-established (active) power synchronization control (PSC) from~\cite{zhang2009power} into vector current control (VCC) that has inherent current limiting capability. It also offers greater flexibility with a point-of-common-coupling (PCC) voltage alignment of the voltage and power references. Additionally, it ensures a smooth transition from more conventional converter control schemes, e.g., those using a PLL, and it is compatible with the generic PLL-based framework of~\cite{harnefors2021generic}. This paper introduces the necessary voltage control modifications that enable the use of universal PSC (UPSC) in the DRU-based OWF setup.

Despite these modifications, the main challenges associated with grid-forming control, stemming from the limited capabilities of the converter and the source, persist.
Significant communication delays between WTs during black starts can cause converters to hit their current or modulation limits. Similarly, a delayed power ramp might necessitate power absorption from some WTs to remain synchronized, which may not be desirable or even feasible for the WT conversion systems. 
Deviating from the original current reference can lead to a loss of synchronism with the grid, characterized by unstable frequency and/or voltage dynamics. 
This paper develops virtual active and reactive power-based control loops to address this issue. These virtual power quantities are calculated using the alternating voltage controller's current references rather than the actual current measurements, ensuring that the outer loops remain unaffected by any subsequent changes to the current references. Virtual active power for the synchronization loop has already been presented by~\cite{kkuni2024effects,laba2023virtual}. During fault clearances and phase jumps, these works have shown that using it can improve large signal stability of the synchronization loop. For the UPSC under consideration, we will demonstrate that both virtual active power and virtual reactive power are essential to enable flawless black start and power ramp stages in OWFs.

Our contributions are as follows. We introduce a UPSC scheme with voltage-power droop controllers in the outer loop to enable its use in OWFs with DRU-based HVDC link. This controller provides a direct coupling between active power and the voltage magnitude. To circumvent stability issues associated with delays in black start and power ramp stages, we introduce virtual active and virtual reactive power. To the best of our knowledge, this is the first publication to employ virtual power quantities within voltage-power droop controllers and also to demonstrate their advantages. Moreover, it is the first publication to highlight the specific benefits of virtual reactive power. Case studies are presented to showcase the performance of the proposed control scheme.

\textit{Notation:}
Complex space vectors in the (synchronous) $\dq$-frame are denoted by boldface letters, e.g.,~$\cxi$. 
When they refer to the (stationary) $\ab$-frame, the superscript~$^s$ is added,~$\cxi^s$. 
Real transfer functions that map between complex space vector spaces---implying no cross-coupling between the two components---are denoted by italic letters, e.g., $G(s)$.  Here, the Laplace variable $s$ is to be interpreted as the derivative operator $s=\frac{d}{dt}$, where appropriate. In the remainder, all quantities (including frequency and time---except in the simulations) are normalized. A first-order low pass filter is denoted by $H_{\alpha}(s)=\frac{\alpha}{s+\alpha}$, and $\alpha$ is referred to as its bandwidth. 
The superscript~$^*$ denotes the (Hermitian) conjugate.

\section{Preliminaries and System Configuration}

\begin{figure}[t!]
	\centering
	\scalebox{0.98}{\begin{tikzpicture}[every text node part/.style={align=center}]
			\node[draw=none,fill=none] at (0,0){\includegraphics[scale=0.66]{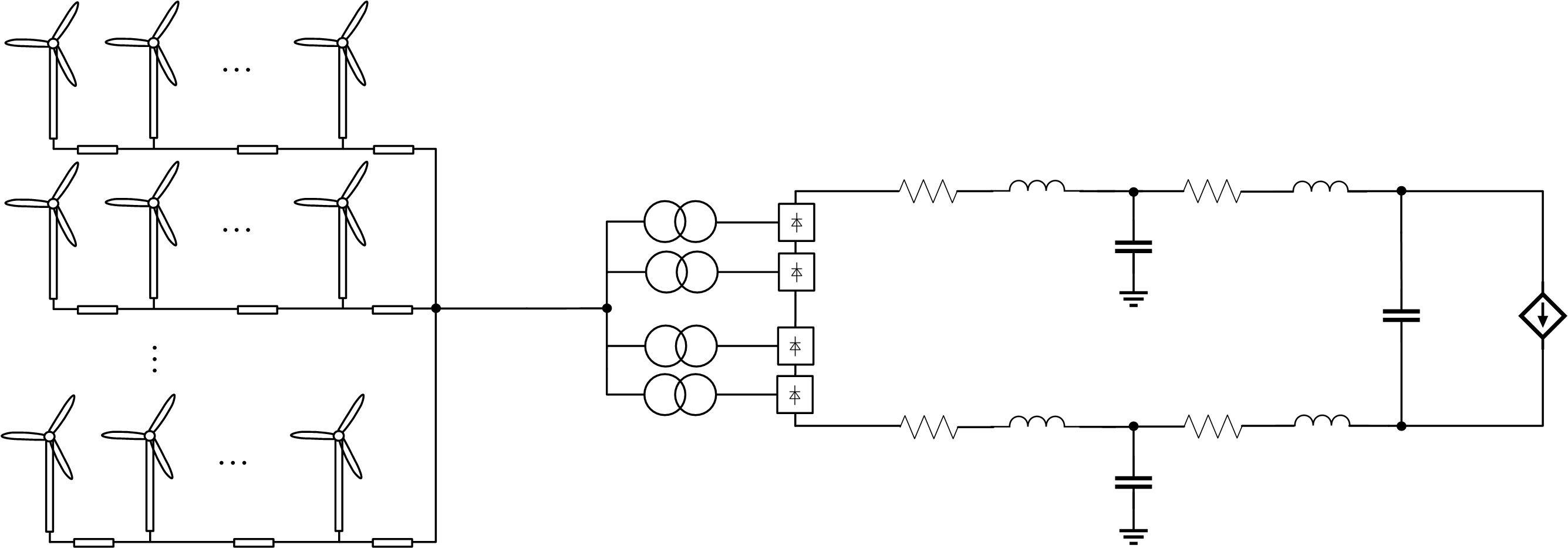}};
			\node[draw=none,fill=none] at (-3.3,1.8) {\scriptsize{Offshore wind farm}};
			\node[draw=none,fill=none] at (-0.4,.9) {\scriptsize{24-pulse DRU}};
			\node[draw=none,fill=none] at (2,.9) {\scriptsize{HVDC link}};
			\node[draw=none,fill=none] at (3.9,-1.2) {$\substack{\text{Onshore}\\ \text{converter}}$};
			\node[draw=none,fill=none] at (-1.65,0) { $\cxv^s_\mathrm{off}$};
			\node[draw=none,fill=none] at (3.6,.7) { $v_\mathrm{on}$};
			\draw[dash pattern=on 1.5pt off 1.5pt] (-4.575,-1.65) rectangle ++(.575,1.05);
	\end{tikzpicture}}
	\caption{OWF with a DRU-based HVDC link.}
	\label{fig:offshore_wind_system}
\end{figure}
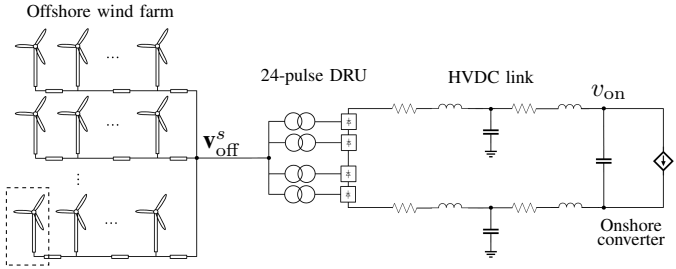

\subsection{OWF configuration}

In Figure~\ref{fig:offshore_wind_system}, an OWF, consisting of multiple strings, is connected to the onshore grid via an HVDC link using a 24-pulse DRU on the offshore side. Optionally, a reactive power compensation unit and/or harmonic filters can be available at the offshore bus, denoted by voltage $\cxv^s_\mathrm{off}$. The controlled current source depicted at the onshore side of the HVDC link approximates the onshore converter. This current source controls $v_\mathrm{on}$, the DC voltage of the HVDC link at the onshore side, using a PI controller and a feedforward. In simulations, this source is restricted from energizing the HVDC link to ensure that the black start stage is entirely managed by the WTs.\looseness=-1


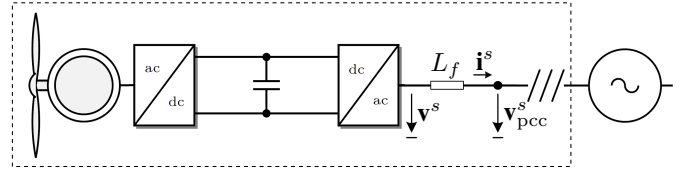
\begin{figure}[t]
	\centering
	\centering
	\scalebox{0.99}{\begin{tikzpicture}[every text node part/.style={align=center}]
			\node[draw=none,fill=none] at (0,0){\includegraphics[width=0.99\columnwidth]{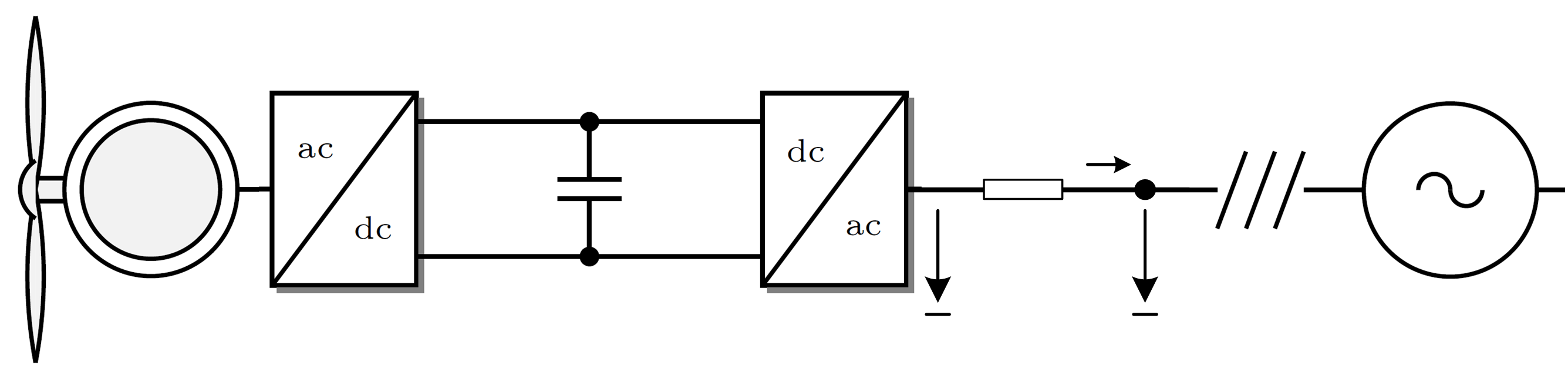}};
			\node[draw=none,fill=none] at (1.325,.25) {$L_f$};
			\node[draw=none,fill=none] at (1.1,-.37) {$\cxv^s$};
			\node[draw=none,fill=none] at (1.85,0.325) {$\cxi^s$};
			\node[draw=none,fill=none] at (2.4,-.425) {$\cxv_\pcc^s$};
			\draw[dash pattern=on 1.5pt off 1.5pt] (-4.5,-1.1) rectangle ++(7.5,2.2);
	\end{tikzpicture}}
	\caption{Example circuit for the WT conversion system.}
	\label{fig:system-model}
\end{figure}

\subsection{Wind converter system}\label{sec:convsys}
Figure~\ref{fig:system-model} illustrates the circuit of the medium-voltage (MV) WT conversion system and the related notation. The conversion system comprises a dual conversion line, as the permanent magnet synchronous generator has two sets
of three-phase windings. In the interest of space, the intricacies of this dual conversion setup are not presented; see~\cite{karaca2024damping} for further details.
On the converter-side of the PCC, we have the stray-inductance of the transformer,~$L_f$. The resistances are small, and they are not presented, as their inclusion in the control scheme is trivial. 
A stiff voltage is illustrated here for presentation purposes only. 
The wind turbine generator-side converters, which are not shown explicitly, are primarily responsible for controlling the DC-link voltages of the WT conversion system. The details of the generator-side control scheme are also omitted in the interest of space.\looseness=-1

\section{Universal Power Synchronization Controller}

\begin{figure*}[t]
	\centering
	\centering
	\scalebox{0.99}{{\begin{tikzpicture}[every text node part/.style={align=center}]
			\node[draw=none,fill=none] at (0,0){\includegraphics[width=1.55\columnwidth]{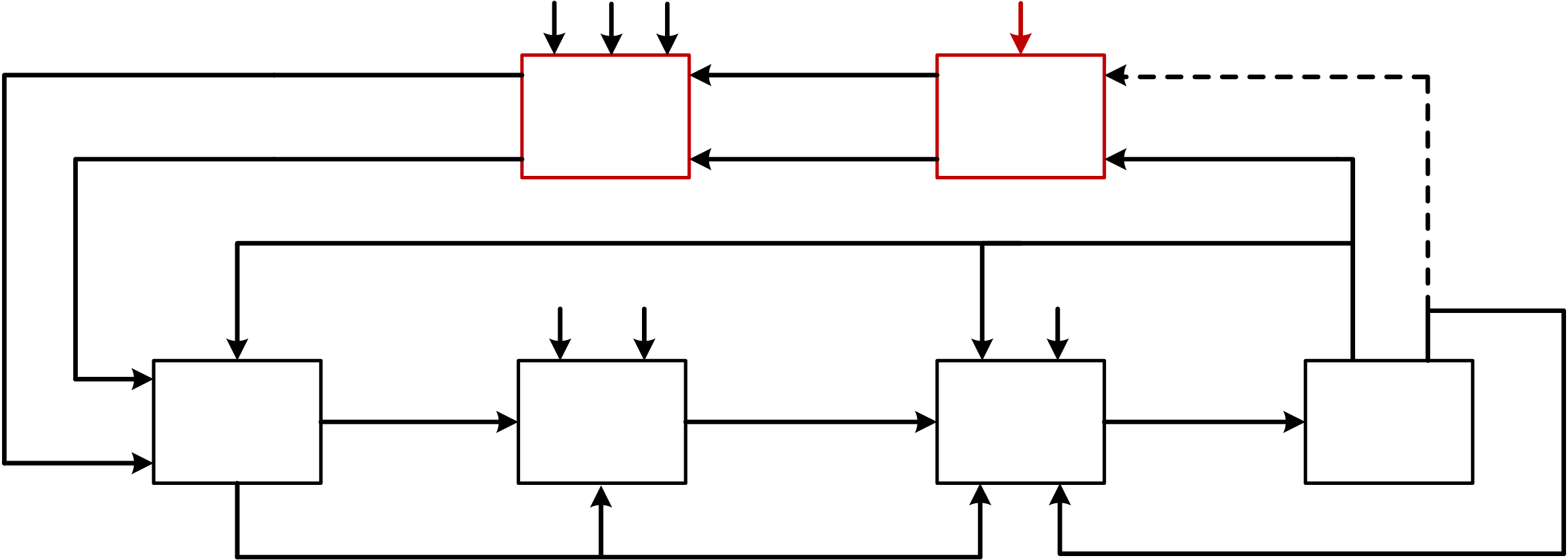}};;
			\node[draw=none,fill=none] at (3.5,2.1) {$\cxi^s$};
			\node[draw=none,fill=none] at (3.5,1.3) {$\cxv_\pcc^s$};
			\node[draw=none,fill=none] at (0.1,2.1) {$\bar P$};
			\node[draw=none,fill=none] at (0.1,1.3) {$\bar Q$};
			\node[draw=none,fill=none] at (-4,2.1) {$\phi$};
			\node[draw=none,fill=none] at (-4,1.3) {$V_\rf$};
			\node[draw=none,fill=none] at (-1.6,2.75) {$P_\rf$};
			\node[draw=none,fill=none] at (-.85,2.75) {$Q_\rf$};
			\node[draw=none,fill=none] at (-2.3,2.75) {$V_\ext$};
			\node[draw=none,fill=none] at (2,2.75) {$\cxi_{\rf0}^s$};
			\node[draw=none,fill=none] at (-1,0) {$I_\mathrm{max}$};
			\node[draw=none,fill=none] at (-2,0) {$P_\mathrm{min}$};
			\node[draw=none,fill=none] at (2.4,0) {$\phi$};
			\node[draw=none,fill=none] at (-3.3,-1) {$\cxi_{\rf0}$};
			\node[draw=none,fill=none] at (0.1,-1) {$\cxi_{\rf}$};
			\node[draw=none,fill=none] at (3.45,-1.05) {$\bm{u}$};
			\node[draw=none,fill=none] at (-5.25,-2.25) {$\cxv_{\pcc}$};
			\node[draw=none,fill=none] at (2.075,1.45) {$\substack{\text{Virtual power}\\ \text{computation}\\ \text{\S\ref{sec:powcom}}}$};
			\node[draw=none,fill=none] at (-4.85,-1.25) {$\substack{\text{Current}\\ \text{reference}\\ \text{\S\ref{sec:curref}}}$};
			\node[draw=none,fill=none] at (-1.6,1.45) {$\substack{\text{PCC}\\ \text{reference}\\ \text{\S\ref{sec::pccref}}}$};
			\node[draw=none,fill=none] at (2.05,-1.25) {$\substack{\text{Current}\\ \text{control}\\ \text{\S\ref{sec:cc}}}$};
			\node[draw=none,fill=none] at (-1.6,-1.25) {$\substack{\text{Current}\\ \text{modification}\\ \text{\S\ref{sec:refmod}}}$};
			\node[draw=none,fill=none] at (5.3,-1.25) {$\substack{\text{Converter}\\ \text{system}\\ \text{\S\ref{sec:convsys}}}$};
	\end{tikzpicture}}}
	\caption{A high-level block diagram of the complete UPSC scheme for grid-side control of the WT conversion system. The red blocks highlight the modifications for the DRU application compared to~\cite{harnefors2020universal}. }
	\label{fig:control_scheme}

\end{figure*}
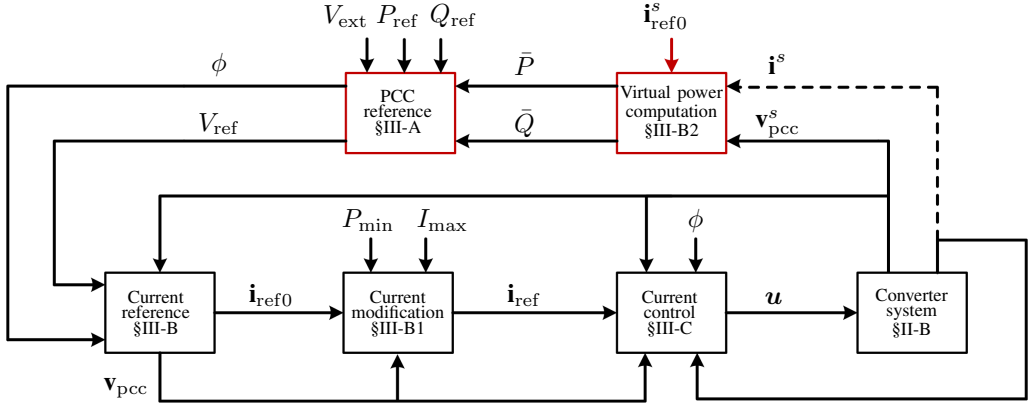

Several definitions are introduced in the following for the complete UPSC scheme provided in Figure~\ref{fig:control_scheme}. The active and reactive power references are denoted by $P_\rf$ and $Q_\rf$, respectively. The active power reference is provided by the high-level wind turbine controller, while the reactive power reference is communicated by the grid operator. An external voltage magnitude reference $V_\ext$ could be communicated or generated by other high-level control loops. The term $P_\mathrm{min}$ denotes the minimum allowed power to limit the reverse power flow, which is nominally set to $0$. The term $I_\mathrm{max}$ is the maximum allowed phase current magnitude. Finally, $\bm{u}$ denotes the switching signal. Other quantities are either defined later or are clear from the context of power conversion control.\looseness=-1

The virtual power computation stage calculates both the active and reactive power, as well as the virtual active and reactive power. The active power is simply $P=\re\{\cxv^s_\pcc\cxi^{s*}\}$, whereas the reactive power is $Q=\im\{\cxv^s_\pcc\cxi^{s*}\}$. From this stage, the terms $\bar P$ and $\bar Q$ are passed on to the PCC reference creation stage, and it will select between the virtual power and the actual measurements. The reasoning behind will be explained in the following sections.
The PCC reference creation stage includes the synchronization loops and the droop controllers comparing $\bar P$ and $\bar Q$ with their respective references. 

The current reference generation stage calculates the current references that would push the PCC voltage towards its reference. Notice that some blocks have combined features, e.g., the current reference stage calculates also the positive sequence of the PCC voltage, to keep the block diagram simple. The current modification stage is where the limited capabilities of the converter and the source are integrated into the scheme. Finally, the current controller closes the loop for current, and creates the switching signal. For the input admittance analysis of the complete scheme under normal operating conditions, the reader is referred to~\cite{karaca2025ontheinput}.\looseness=-1

\subsection{PCC voltage reference generation}\label{sec::pccref}

\subsubsection{Synchronization and the $\dq$-frame (reference) angle}

The power synchronization loop generates the $\dq$-frame angle~$\phi$. This angle is also the reference angle for the PCC voltage: 
\begin{equation}\label{eq:sync_loop}
		\phi = (1/s)\left[\omega_1 + K_P(s)(P_\rf - \bar P)\right],
\end{equation}
where $\omega_1$ is the nominal synchronous frequency, and
$K_P(s) = \frac{sT_d+1}{sM+k_m}$,
with $T_d$ being the time constant used for damper winding emulation, $k_m$ being the frequency droop constant, and $M$ being the virtual inertia constant. 

\subsubsection{Voltage magnitude reference}

The PCC voltage magnitude reference is generated by
\begin{equation}\label{eq:droop}
	\begin{split}
		V_\rf = V_\ext &+ K_{QV}(s) \left[Q_\rf - H_{\alpha_Q}(s)\bar Q\right]\\ &+ K_{PV}(s) \left[P_\rf - H_{\alpha_P}(s) \bar P\right].
	\end{split}
\end{equation}
This block includes:

\quad \textit{(a) }\textit{$QV$ control/droop:} A $QV$ droop functionality, where $K_{QV}(s)=K_{QV}$ and $\alpha_Q<\omega_1$, is necessary for reactive power sharing in parallel operation~\cite{chandorkar2002control}. During black start scenarios with communication delays, $QV$ droop synchronizes the voltage magnitude ramps across multiple converters, see Section~\ref{sec:blackstart}. 
	
\quad \textit{(b) }\textit{$PV$ control/droop:} A $PV$ control functionality, where $K_{PV}(s)=K_{PV}+\frac{K_{PV,I}}{s}$ and $\alpha_P<\omega_1$, is essential to push power over loads with strong voltage sensitivity, such as DRU.
Specifically, the active power flow is proportional to the voltage magnitude (see the models provided in~\cite{bidadfar2020control,tang2021dru}). Thus, this is the primary component that enables the use of UPSC in the DRU-based OWF setup. Moreover, $PV$ droop, the proportional part of $PV$ control, can improve passivity in the low frequency range~\cite[{\S}IV]{karaca2025ontheinput}. The tuning here must be based on the dynamical requirements set for the OWF. During power ramp scenarios with delays, $PV$ control also ensures a synchronous voltage increase, and thus maintains a balanced reactive power sharing among WTs, see Section~\ref{sec:powerramp}. \looseness=-1

\subsection{Current reference generation}\label{sec:curref}
The alternating voltage controller (AVC) generates:
{\medmuskip=.7432mu \thickmuskip=.7432mu \thinmuskip=.7432mu\begin{equation}\label{eq:psavc}
	\cxi_{\rf0}= \frac{P_\rf-jQ_\rf}{V_\rf}+\frac{1}{R_a}\left( 1 + \frac{\alpha_a}{s}\right) \left[V_\rf-{H}_{\alpha_{F}}(s)\cxv_{\pcc}\right],
\end{equation}}where $R_a$ is the proportional gain of the inner current controller, $\alpha_a$ is the integral control bandwidth (generally chosen small $\alpha_a<0.05\omega_1$), and $\alpha_{F}$ is the PCC voltage feedforward bandwidth (e.g., $\alpha_{F}<R_a/L_f$). The first additive term is the reference feedforward proposed in~\cite{harnefors2020reference}, while the second additive term, derived in~\cite[{\S}III]{harnefors2020universal}, unifies VCC and PSC.
These current references can also be mapped to the $\ab$-frame as $\cxi_{\rf0}^s$ using the angle $\phi$.\looseness=-1
\subsubsection{Reference modifications}\label{sec:refmod}
Two critical considerations are listed for the limitations of the converter and the source.\looseness=-1

\quad \textit{(a) }\textit{Reverse power limitation:} The following equation limits the reverse power flow:
	\begin{equation}
		\cxi_{\rf \rr} =\cxi_{\rf0}-  \frac{\cxv_{\pcc,f}}{|\cxv_{\pcc,f}|^2} \min\{0,\re\{\cxv_{\pcc,f}\cxi_{\rf0}^*\}-P_\mathrm{min}\},
\end{equation}
where $\cxv_{\pcc,f}={H}_{\alpha_{F}}(s)\cxv_{\pcc}$. In essence, if the AVC current reference, $\cxi_{\rf0}$, would result in a reverse power exceeding the threshold $P_\mathrm{min}$, a simple projection step removes the excess while preserving the reactive power component specified by the AVC.

\quad \textit{(b) }\textit{Current magnitude limitation:}  Converters can generally tolerate a maximum current of approximately 120\% of the rated current for extended periods~\cite{BoFan2022}. During transients, however, the AVC may generate a reference that exceeds this limit, denoted as $I_\mathrm{max}$. Angle preserving scaling is used for magnitude limitation ensuring that the reverse power limitation is also preserved:\looseness=-1
	\begin{equation}
	\cxi_{\rf} =\dfrac{I_\mathrm{max}}{\max\{|\cxi_{\rf \rr}|,I_\mathrm{max}\}}\cxi_{\rf \rr}.
\end{equation}

\begin{figure}[t]
	\centering
	\begin{tikzpicture}[semic/.style args={#1,#2}{semicircle,minimum width=#1,draw,anchor=arc end,rotate=#2},outer sep=0pt,line width=.7pt]
		\pgfmathsetmacro{\ATAN}{atan{-9/4}}
		\draw[<->] (-2,0) -- (2,0) node[anchor=west] {$\scriptstyle\dcom$};
		\draw[<->] (0,-2) -- (0,2) node[anchor=south] {$\scriptstyle\qcom$};
		\draw[thick] (0,0) circle (1.2cm);
		\draw[-] (-8/9,2) -- (8/9,-2);
		\draw[->] (0,0) -- (.675,.3);
		\coordinate (A) at (0.225,0.1);
		\coordinate (B) at (0,0);
		\coordinate (C) at (-0.1,0.225);
		\draw[-] (A) -- (0.115,0.325);
		\draw[-] (0.115,0.325) -- (C);
		\node [semic={2.4cm,\ATAN}, black, fill=gray, fill opacity=0.2] at (-0.48736,1.09656){};
		\node (A2) at (0.57,0.53) {$\cxv_{\pcc,f}$};
		\node (A3) at (0.4,1.44) {$I_\mathrm{max}$};
		\node (A4) at (-1.75,0.3) {$-I_\mathrm{max}$};
		\node (A5) at (1.65,-0.3) {$I_\mathrm{max}$};
		\node (A6) at (-0.55,-1.44) {$-I_\mathrm{max}$};
		\node (A7) at (0.0575,0.1625) {$\cdot$};
	\end{tikzpicture}
	\caption{Example for the current limits when $P_\mathrm{min}=0$. The shaded area represents the safe operating region.}\label{fig:const}
\end{figure}
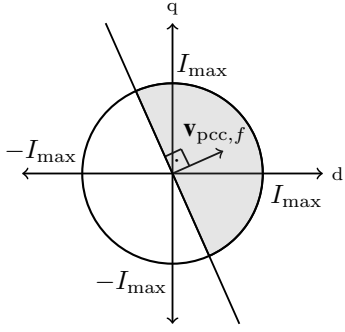

Figure~\ref{fig:const} illustrates an example constraint set for $P_\mathrm{min}=0$. The two reference modification steps above sequentially constrain the current reference $\cxi_{\rf0}$ within the safe operating region, which is depicted by the shaded area.

\subsubsection{Virtual power computation}\label{sec:powcom}
Whenever any of the modifications above are active, or whenever the current controller itself does not exhibit a good tracking performance (e.g., due to modulation index limits), we face the risk of loss of power sharing or even loss of synchronism. To this end, the virtual active and reactive power terms are defined and calculated based on the unmodified AVC current references:
\begin{equation}
		P_\vir=\re\{\cxv^s_\pcc\cxi_{\rf0}^{s*}\},\quad
		Q_\vir=	\im\{\cxv^s_\pcc\cxi_{\rf0}^{s*}\}.
\end{equation}
For the control scheme presented in the case studies, we replace the measurements with their virtual counterparts, i.e., $\bar P=P_\vir$ and $\bar Q=Q_\vir$, and this substitution is active at all times. In other words, the synchronization loop in~\eqref{eq:sync_loop} and the $QV$ and $PV$ controllers in~\eqref{eq:droop} operate exclusively with virtual active and reactive power, ensuring that their behavior remains unaffected by any subsequent changes in the currents. This design choice demonstrates better the effectiveness of virtual power, and emphasizes that no detection mechanism is required to activate the virtual power-based control loops. The corresponding block in Figure~\ref{fig:control_scheme} still takes the actual current measurements as an optional (dashed) input, allowing for the implementation of a switch-over logic for this selection process if desired. 
\subsection{Current control design}\label{sec:cc}
The control scheme described up to now is compatible with various current controllers. We employ an $\ab$-frame current controller with a PCC voltage feedforward which also accounts for the voltage drop across the inductance $L_{f}$:
\begin{equation}\label{eq:modcc}
	\cxv^s_\rf = R_a(\cxi_{\rf}^s-\cxi^s)+ j\omega_1L_f\cxi_{\rf}^s+\cxv_{\pcc,f}^s,
\end{equation}
where $R_a$ is the proportional gain, and $\cxv^s_\rf$ is the reference voltage (e.g., to be created via PWM).
For MV applications, model predictive pulse pattern control (\mpppc) could be considered essential, since optimized pulse patterns bring in significant advantages, and \mpppc\ maximizes the benefits of them~\cite{geyer2011model}. As a remark, the controller above has been shown to approximate \mpppc~well, see~\cite{karaca2024damping}.\looseness=-1
\section{Case Studies}
\begin{figure}[t!]
	\centering
	{\begin{tikzpicture}[every text node part/.style={align=center}]
			\node[draw=none,fill=none] at (0,0){\includegraphics[scale=0.66]{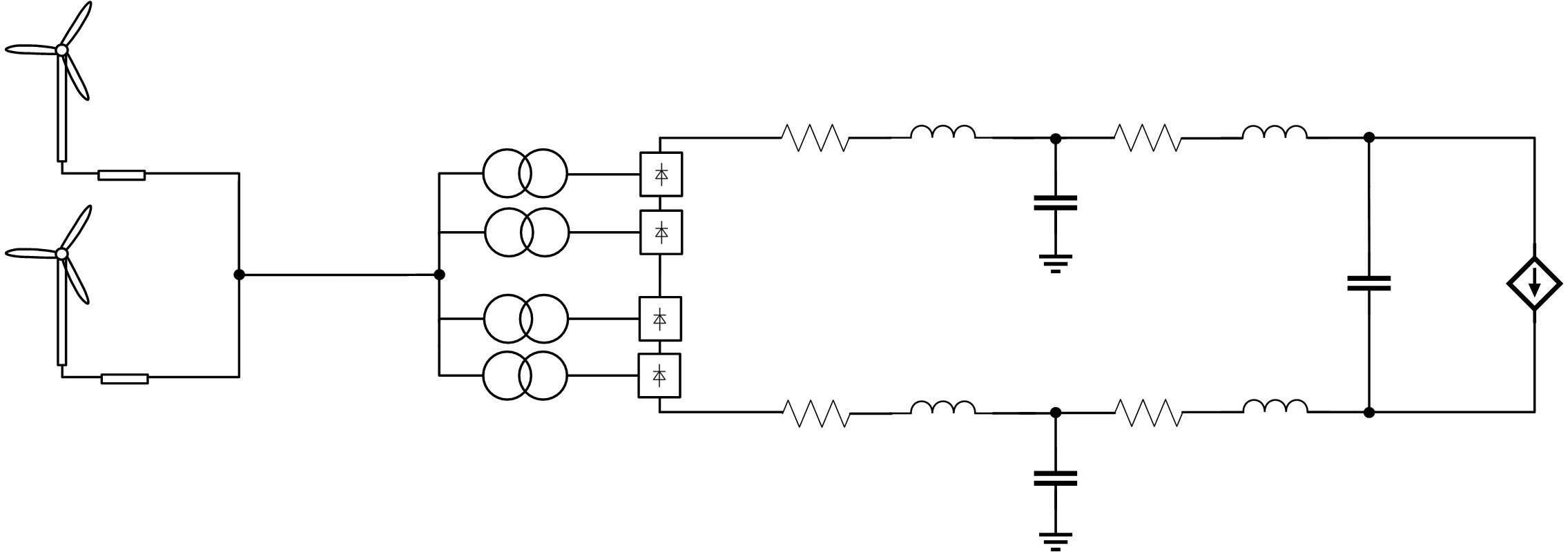}};
			\node[draw=none,fill=none] at (-4.1,.65) {$\substack{\text{WTS1:}\\ 36\, \text{WTs}}$};
			\node[draw=none,fill=none] at (-4.1,-.35) {$\substack{\text{WTS2:}\\ 38\, \text{WTs}}$};
	\end{tikzpicture}}
	\caption{Simulation environment with string aggregation.}
\label{fig:offshore_wind_system2}
\end{figure}

We conduct studies in an environment---implemented in MATLAB Simulink---where the dual converter system of~\cite{karaca2024damping} is used as an interface for a $19.1\,\text{MVA}$ WT generator. The setup, illustrated in Figure~\ref{fig:offshore_wind_system2}, consists of two aggregated strings. WTS1, the top string, is an aggregation of $36$ WTs, whereas WTS2, the bottom string, is an aggregation of $38$ WTs. The relevant system parameters and rated values are given in Table~\ref{tab:rated values}. Other parameters (e.g., those relating to the cables, diode rectifier, its transformer, and the HVDC link) are considered confidential. The base systems are defined with respect to the aggregations.\looseness=-1

The controlled current source on the onshore side regulates the DC voltage with a control and feedforward bandwidth of $25$ Hz. As mentioned before, this source is restricted from black starting the HVDC link. The control parameters are detailed in Table~\ref{tab:param}. Since we are interested in system behavior in the low-frequency range, we work with averaged converter models for the WT conversion system. However, the switching and control system delays are taken into account, e.g., by considering the actual sampling intervals and control time scales of our converters. To further simplify the setup, we assume that the reactive power reference is not communicated (i.e., $Q_\rf=0$), and $V_\ext$ is generated locally, meaning there is no secondary supervisory control. To abide by the page limitations, we present only the most relevant plots. \looseness=-1

\begin{table}[t!]
	\caption{WT system parameters and rated values.}
	\label{tab:rated values}
	\centering
	\begin{tabular}{lcl}
		\hline
		Parameter & Symbol & Value \\
		\hline
		Rated apparent power of the transformer & $S_{\mathrm{R},\mathrm{trafo}}$ & $\SI{18}{{\mega\voltampere}}$ \\ 
		Rated voltage at the primary (grid) & $V_{\mathrm{R},\mathrm{prim}}$ & $\SI{66}{\kilo\volt}$ \\
		Rated voltage at the secondary (converter) & $V_{\mathrm{R},\mathrm{sec}}$ & $\SI{3.1}{\kilo\volt}$ \\
		Rated angular frequency &  $\omega_1$ & $\SI[parse-numbers=false]{2 \pi 50}{\radian\per\second}$ \\
		Filtering inductance equivalent for trafo & $L_f$ & $\SI{0.18}{\pu}$\\
		Filtering resistance equivalent for trafo & $R_f$  & $\SI{0.01}{\pu}$\\
		DC-link voltage of WT & $V_\dc$  & $\SI{1.9754}{\pu}$ \\
		\hline
	\end{tabular}
\end{table}

\begin{table}[!t] 
	\renewcommand{\arraystretch}{1}
	\caption{Control parameters in per unit.}
	\label{tab:param}
	\noindent
	\centering
	\begin{minipage}{\linewidth}
		\begin{center}
			\begin{tabular}{lc|lc}
				\hline
				Parameter/Symbol & Value & Parameter/Symbol & Value\\
				\hline
				$H$ (Inertia time const.) & $\SI{1}{{s}}$ & $R_a$  & $\SI{0.36}{{\pu}}$ \\ 
				$k_m$ & $\SI{20}{{\pu}}$ &	$\alpha_a$ & $\SI{0.01}{{\pu}}$ \\
				$T_d$ & $\SI{0}{{\pu}}$ & $\alpha_F$ & $\SI{2}{{\pu}}$ \\
				$K_{QV}$ & $\SI{0.05}{{\pu}}$ &	$I_\mathrm{max}$ & $\SI{1.2}{{\pu}}$ \\
				$\alpha_Q$ & $\SI{0.2}{{\pu}}$ & $K_{PV}$ & $\SI{0.75}{{\pu}}$\\ 
				$\alpha_P$ & $\SI{0.5}{{\pu}}$&$K_{PV,I}$ & $\SI{5}{{\pu}}$\\ 
				\hline
			\end{tabular}
		\end{center}
	\end{minipage}
\end{table}


\subsection{Black start with communication delays}\label{sec:blackstart}
In this case study, we analyze the black start procedure that is performed by the two strings.
Their voltage ramps are set to $\SI{0.8}{{\pu}}$ with slope $\SI{0.6}{{\pu/s}}$. WTS2 receives the voltage ramp start signal with a communication delay of $\SI{300}{{ms}}$, which is large compared to typical communication delays in this setup, e.g., $50$ to $\SI{100}{{ms}}$. Whenever virtual power is disabled for a control loop, it means that the actual power measurements are used instead.\looseness=-1

\subsubsection{Virtual power disabled only in $QV$ and $PV$ control, but enabled in the synchronization loop}

\begin{figure}[t]
		\centering
	\begin{subfigure}{0.49\columnwidth}
		\centering
		\includegraphics[width=1\columnwidth]{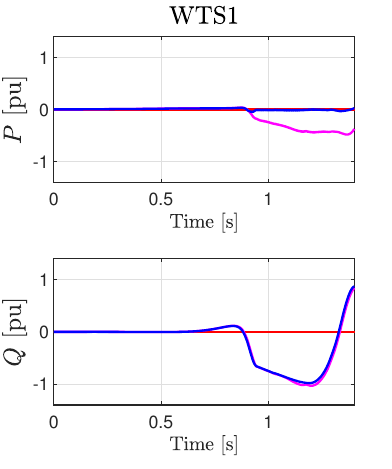}
	\end{subfigure}
	\begin{subfigure}{0.49\columnwidth}
		\centering
		\includegraphics[width=1\columnwidth]{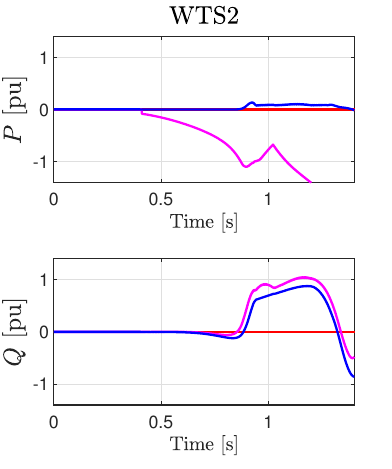}
	\end{subfigure}
	\caption{Delayed black start: Virtual power disabled. \textit{(Red: Reference. Blue: Measurement. Purple: Virtual)}}\label{fig:A1}
\end{figure}

Figure~\ref{fig:A1} illustrates the loss of synchronism that occurs when WTS2 joins the black start late, and if the $QV$ and $PV$ control loops do not utilize virtual power. This happens because both strings reach their current and reverse power limits due to the delay between the voltage ramps. The voltage-power control loops attempt to correct this while being unaware of the limits. Similar observations can be made if one of the virtual powers in the $QV$ or $PV$ control is disabled.\looseness=-1
\subsubsection{Virtual power enabled in all controllers}

\begin{figure}[t]
		\centering
	\begin{subfigure}{0.49\columnwidth}
		\centering
		\includegraphics[width=1\columnwidth]{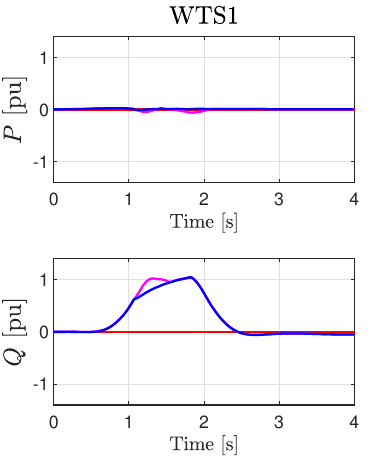}
	\end{subfigure}
	\begin{subfigure}{0.49\columnwidth}
		\centering
		\includegraphics[width=1\columnwidth]{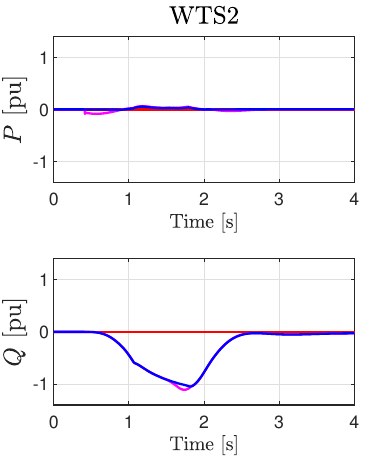}
	\end{subfigure}
	\begin{subfigure}{1\columnwidth}
		\centering
		\includegraphics[width=1\columnwidth]{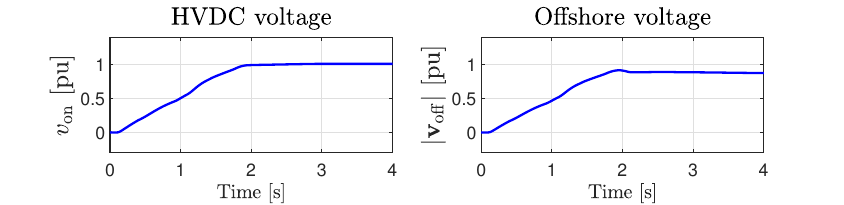}
	\end{subfigure}
	\caption{Delayed black start: Virtual power enabled.}\label{fig:A2}
\end{figure}

Figure~\ref{fig:A2} showcases the flawless black start under excessive communication delays despite having no secondary supervisory control. One can observe the time instances where we benefit from having virtual power-based control loops especially from the difference between the virtual reactive power and the reactive power measurements in the plots.\looseness=-1

\subsection{Delayed power ramps}\label{sec:powerramp}
Once the black start is completed, the two strings enter the power ramp stage. In this scenario, WTS2 ramps with a delay of $\SI{1}{{s}}$, when compared to WTS1.

\subsubsection{Virtual power disabled in all control loops and the limit $P_\mathrm{min}=-\infty$}

\begin{figure}[t]
		\centering
	\begin{subfigure}{0.49\columnwidth}
		\centering
		\includegraphics[width=1\columnwidth]{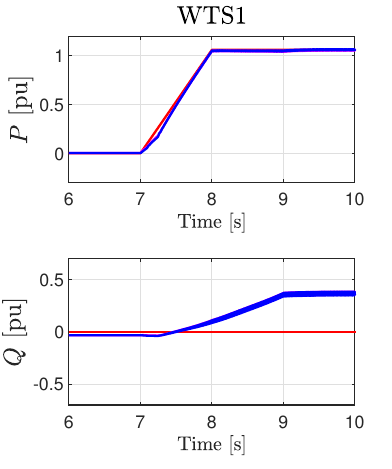}
	\end{subfigure}
	\begin{subfigure}{0.49\columnwidth}
		\centering
		\includegraphics[width=1\columnwidth]{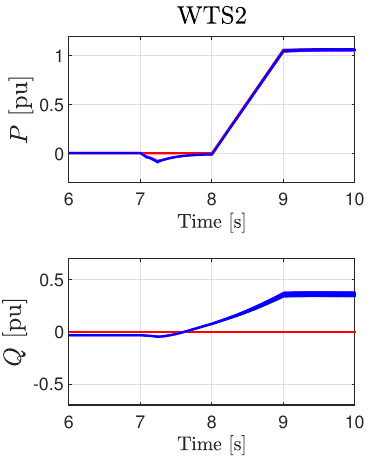}
	\end{subfigure}
	\caption{Delayed ramp: Virtual power disabled and $P_\mathrm{min}=-\infty$.}\label{fig:B1}
\end{figure}

Figure~\ref{fig:B1} illustrates that when reverse power limits are removed, the use of virtual power is no longer necessary to handle delayed power ramps. The $PV$ control successfully achieves a synchronous voltage increase, which can be confirmed from the balanced reactive power sharing.\looseness=-1

\subsubsection{Virtual power disabled only in $PV$ control and the limit $P_\mathrm{min}=0$}

\begin{figure}[t]
		\centering
	\begin{subfigure}{0.49\columnwidth}
	\centering
	\includegraphics[width=1\columnwidth]{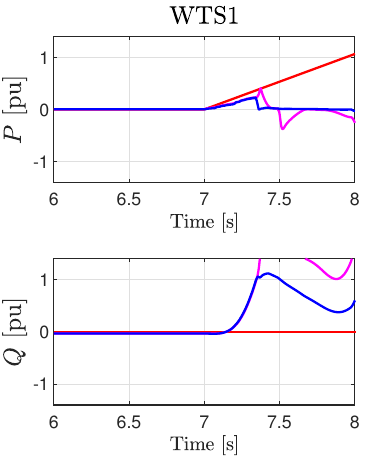}
\end{subfigure}
\begin{subfigure}{0.49\columnwidth}
	\centering
	\includegraphics[width=1\columnwidth]{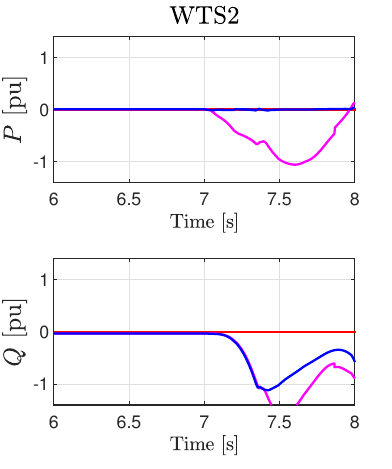}
\end{subfigure}
	\caption{Delayed ramp: Virtual power disabled and $P_\mathrm{min}=0$.}\label{fig:B2}
\end{figure}
Figure~\ref{fig:B2} demonstrates that the absence of virtual power in $PV$ control alone can prevent the power ramp. During asynchronous power ramps, it is imperative for WTS1 and WTS2 to increase voltage synchronously. Failure to do so results in a large reactive power transfer from WTS1 to WTS2, causing WTS1 to reach its current limits. Observations related to loss of synchronism can be made if instead the virtual power in power synchronization loop is disabled.\looseness=-1

\subsubsection{All virtual powers enabled and with $P_\mathrm{min}=0$}

\begin{figure}[t]
	\centering
		\begin{subfigure}{0.49\columnwidth}
		\centering
		\includegraphics[width=1\columnwidth]{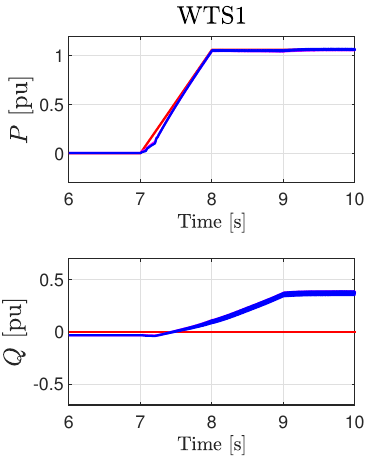}
	\end{subfigure}
	\begin{subfigure}{0.49\columnwidth}
		\centering
		\includegraphics[width=1\columnwidth]{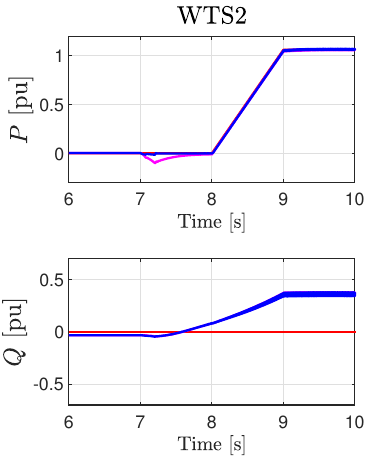}
	\end{subfigure}
	\begin{center}
	\begin{subfigure}{1\columnwidth}
		\centering
		\includegraphics[width=1\columnwidth]{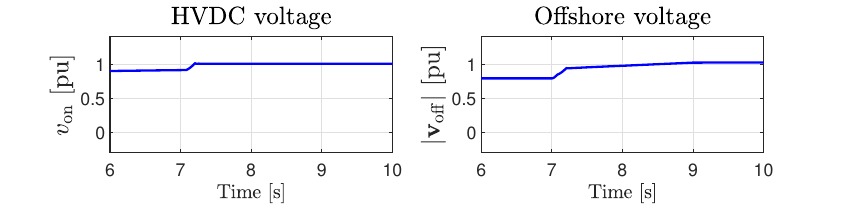}
	\end{subfigure}
	\end{center}
	\caption{Delayed ramp: Virtual power enabled and $P_\mathrm{min}=0$.}\label{fig:B3}
\end{figure}

Figure~\ref{fig:B3}  showcases the flawless power ramp. The virtual reverse power absorbed by WTS2 ensures synchronous voltage increase whenever necessary.\looseness=-1

\section{Conclusion}
We presented a UPSC scheme that enables flawless black start and power ramp stages in OWFs with a DRU based HVDC link connection. To achieve this, a voltage magnitude reference was generated by $QV$ and $PV$ voltage-power droop controllers. To address the stability issues arising from the limited capabilities of the converter and the source, active and reactive power measurements were replaced by their virtual counterparts using the original (unmodified) current references generated by the AVC. 

To the best of our knowledge, this is the first publication to employ virtual power quantities within $QV$ and $PV$ control loops and also to demonstrate the advantages to do so. Moreover, it is the first publication to highlight the specific benefits of virtual reactive power. Case studies confirmed the improved robustness of the controller for the DRU-based OWF setups considered in the $\text{FlexH}_2$ project. 

\bibliographystyle{IEEEtran}
\bibliography{virtual_report}

\end{document}